\documentclass[12pt]{iopart}

\usepackage{iopams}
\usepackage{amssymb}
\usepackage{setstack}
\usepackage{graphicx}
\usepackage{bm}
\usepackage{color}

\begin{document}

\title{Nonequilibrium Kondo transport through a quantum dot in a magnetic field}

\author{Sergey Smirnov and Milena Grifoni}
\address{Institut f\"ur Theoretische Physik, Universit\"at Regensburg, D-93040 Regensburg, Germany}
\ead{Sergey.Smirnov@physik.uni-regensburg.de, Milena.Grifoni@physik.uni-regensburg.de}

\date{\today}

\begin{abstract}
We analyze universal transport properties of a strongly interacting quantum dot in the Kondo regime when
the quantum dot is placed in an external magnetic field. The quantum dot is described by the asymmetric
Anderson model with the spin degeneracy removed by the magnetic field resulting in the Zeeman splitting.
Using an analytical expression for the tunneling density of states found from a Keldysh effective field
theory, we obtain in the whole energy range the universal differential conductance and analytically
demonstrate its Fermi-liquid and logarithmic behavior at low- and high-energies, respectively, as a
function of the magnetic field. We also show results on the zero temperature differential conductance as
a function of the bias voltage at different magnetic fields as well as results on finite temperature
effects out of equilibrium and at a finite magnetic field. The modern nonequilibrium experimental issues
of the critical magnetic field, at which the zero bias maximum of the differential conductance starts to
split into two maxima, as well as the distance between these maxima as a function of the magnetic field
are also addressed.
\end{abstract}


\maketitle

\section{Introduction}
Among various quantum many-particle phenomena in condensed matter physics the Kondo effect
\cite{Hewson_1997} stands out as a unique fundamental state exhibiting remarkable properties both in
bulk systems \cite{de_Haas_1934,Kondo_1964} and in a quantum dot (QD) setup
\cite{Glazman_1988,Meir_1993,Ralph_1994,Goldhaber-Gordon_1998}.

In QDs \cite{Reed_1988} the Kondo effect appears when a QD is switched into the Coulomb blockade regime
by tuning the gate voltage. Normally the QD states with even and odd numbers of electrons (Coulomb
valleys) result in a zero bias minimum of the differential conductance as a function of the bias voltage.
However, when the temperature $T$ is lowered this situation changes qualitatively: the QD states with
even numbers of electrons still result in a zero bias minimum of the differential conductance while for
the QD states with odd numbers one observes an anomalous behavior where the differential conductance at
zero bias shows a maximum often referred to as the Kondo resonance. This so-called zero bias anomaly is
a manifestation of the unpaired spin inherent to the QD states with odd numbers of electrons.

Another important property of the Kondo effect is that together with its appearance the zero bias anomaly
acquires universality: the differential conductance of different QDs collapses into a single function of
the temperature and voltage. It depends on specific QD parameters only through the energy scale
$kT_{\rm K}$, where $T_{\rm K}$ is the Kondo temperature.

The Kondo effect has different universal character at low and high energies. At low energies, smaller than
$kT_{\rm K}$, one observes the Fermi-liquid behavior (strong coupling limit), a quadratic universal
dependence of the differential conductance on the temperature, voltage, or external magnetic field. In the
opposite regime of high energies, larger than $kT_{\rm K}$, the differential conductance is characterized
by a logarithmic universal dependence (weak coupling limit). At intermediate energies it shows a universal
crossover between these two limits.

The zero bias anomaly and its universality have been observed in equilibrium \cite{Goldhaber-Gordon_1998a}
and nonequilibrium \cite{Grobis_2008}. The nonequilibrium Kondo universality in the deep crossover is
characterized by a universal value of the differential conductance at the Kondo voltage, $eV=kT_{\rm K}$.
Theory predicts \cite {Pletyukhov_2012,Smirnov_2013} that this universal value is equal to $2/3$ of the
maximum value which the differential conductance takes in equilibrium at zero temperature. This provides an
efficient way to nonequilibrium experimental measurements of the Kondo temperature \cite{Kretinin_2012}.

Interplay between the Kondo state and other quantum collective phenomena such as ferromagnetism
\cite{Gaass_2011}, superconductivity \cite{Martin-Rodero_2011,Kim_2013} or the Kondo state in an external
magnetic field \cite{Quay_2007,Kretinin_2011} currently represent a very active research field both in
theory and experiment.

The Kondo effect in a QD placed in an external magnetic field is of particular interest. The magnetic field
couples to the spins of the electrons in the QD and produces the Zeeman splitting of the single-particle QD
energy level. It is known \cite{Meir_1993,Ralph_1994,Goldhaber-Gordon_1998} that such a magnetic field
destroys the zero bias anomaly, that is, it restores the normal behavior where the QD states with odd
numbers of electrons at low voltages result in minima of the differential conductance, similar to the QD
states with even occupancies.

However, it turns out that the magnetic field does not destroy the universality. Theory \cite{Hewson_1997}
and experiment \cite{Kretinin_2011} demonstrate that even when there appears a minimum in the differential
conductance at $V=0$, it still remains universal both in equilibrium (as a function of $T$ at $V=0$) and
nonequilibrium (as a function of $T$ and $V$).

Important nonequilibrium issues which have been addressed in experiments \cite{Quay_2007,Kretinin_2011} on
the Kondo effect in an external magnetic field are 1) Kondo universality; 2) the critical magnetic field
at which the zero bias maximum of the differential conductance starts to split into two maxima; 3) the
distance between these maxima as a function of the magnetic field; 4) the high-field limit of this distance.
The issues above have obviously a highly nonequilibrium nature. The nonequilibrium theory of the Kondo
effect in the {\it whole} energy range is a complicated problem even without a magnetic field. The numerical
renormalization group (NRG) method \cite{Bulla_2008}, which is known to be a numerically exact tool in
equilibrium, is difficult to generalize to nonequilibrium \cite{Anders_2008,Schmitt_2011}. To cover the whole
energy range in nonequilibrium one has to resort to other methods. These are the non-crossing approximation
(NCA) (intermediate and large energies) \cite{Meir_1993,Wingreen_1994,Aguado_2003}, equations of motion
(mainly qualitative tool) \cite{Meir_1993,Martinek_2003,Fang_2008}, mean-field theories (low energies)
\cite{Aguado_2000,Lopez_2003}, or $1/N$ expansions (low energies) \cite{Ratiani_2009}. In the whole energy
range the real-time renormalization group method \cite{Pletyukhov_2012} is a powerful tool. It has been
developed for the $s-d$ model \cite{Hewson_1997} and can account for the Zeeman splitting in the weak
coupling regime \cite{Korb_2007,Schoeller_2009,Schoeller_2009a}. To our knowledge, however, a generalization
to include the Zeeman splitting in the strong coupling regime has not been done yet. Another theoretical
tool is the Keldysh effective field theory \cite{Smirnov_2011,Smirnov_2011a,Smirnov_2013} developed for the
Anderson model \cite{Anderson_1961}. In its first application it has been shown that this theory well
captures the weak coupling limit \cite{Smirnov_2011} and is capable to cope with finite, though large
interaction strengths \cite{Smirnov_2011a}. Recently it has been further developed to describe the behavior
of the differential conductance in the whole range of temperatures and bias voltages \cite{Smirnov_2013}.

In the present work we address the behavior of the Kondo state in an external magnetic field both in
equilibrium and nonequilibrium. To this end we extend the Keldysh effective field theory \cite{Smirnov_2013}
to account for the Zeeman splitting of the single-particle QD energy level and derive an approximate analytical
expression for the QD tunneling density of states (TDOS). The theory is then applied to calculate the universal
differential conductance as a function of the temperature, bias voltage and magnetic field in the whole energy
range. At this point we would like to note that while there are other theories able to address the Kondo effect,
they are only applicable to specific regimes of the Kondo state. In particular, the perturbative renormalization
group \cite{Rosch_2003a,Rosch_2003,Rosch_2005} is applicable in the weak coupling (or logarithmic) regime,
renormalized perturbation theory \cite{Hewson_2005} or interpolative perturbative approximation
\cite{Aligia_2006} are applicable in the strong coupling (or Fermi-liquid) regime, and NCA \cite{Roura-Bas_2010}
is applicable and very successful in the weak coupling regime and also in the intermediate coupling regime
(several orders below $T_{\rm K}$). This shows that a single theory able to describe all the regimes of the Kondo
state is highly desirable. In the present work we make a step to develop such a single theory and demonstrate
that despite the approximations made in the theoretical derivation our theory has a number of advantages:
1) Fermi-liquid behavior when the Zeeman energy is much less than $kT_{\rm K}$ (advantage over NCA
\cite{Hewson_1997}); 2) logarithmic behavior when the Zeeman energy is much larger than $kT_{\rm K}$ (advantage
over mean-field theories \cite{Hewson_1997}); 3) critical magnetic field close to results of previous theories
\cite{Hewson_2005,Aligia_2006} and experiments \cite{Quay_2007}; 4) distance between the maxima of the
differential conductance as a function of the magnetic field in a good agreement with experiments
\cite{Quay_2007} and absence of a spurious zero bias peak at a finite magnetic field (advantage over NCA
\cite{Meir_1993}); 5) high-field limit of this distance in a good agreement with experiments \cite{Quay_2007};
6) universality with the correct scaling given by $T_{\rm K}$.

The paper is organized as follows. In Section \ref{TM} we present our model of a strongly correlated
QD in a magnetic field. Here we also show the slave-bosonic transformation used in our work. The analytical
expression for the QD TDOS resulting from an effective Keldysh field theory is given in Section \ref{KFIS}.
Here we analytically investigate the low-energy sector of the theory proving its Fermi-liquid nature and
providing the Fermi-liquid coefficients for the temperature, voltage and magnetic field behaviors as well as
their universal ratios. Additionally, when the Zeeman energy is much larger than $kT_{\rm K}$ we analytically
derive the logarithmic asymptotics of the differential conductance as a function of the magnetic field at $T=0$
and $V=0$. In Section \ref{Results}, integrating the analytical expression for the QD TDOS, we calculate the
universal behavior of the differential conductance in the whole range of temperatures, voltages and magnetic
fields. Here we show all the regimes of the Kondo state: the strong coupling limit, weak coupling limit and the
crossover region. Finally, we conclude the paper in Section \ref{Conclusion} where all the advantages and
drawbacks of our theory are summarized and a systematic way to improve the theory is mentioned.
\section{Theoretical model}\label{TM}
To describe the strongly interacting QD we employ the single-impurity Anderson model \cite{Anderson_1961}
(SIAM). This model describes a system of two interacting electrons. The strength of the electron-electron
interaction is given by the parameter $U>0$. In the absence of the interaction the electrons can occupy a spin
degenerate single-particle energy level $\epsilon_{\rm d}$. We include the effect of an external magnetic field
through the Zeeman splitting $\Delta\epsilon$ of this energy level, $\epsilon_{\rm d}\rightarrow\epsilon_\sigma$
with $\epsilon_\sigma\equiv\epsilon_{\rm d}+\sigma\Delta\epsilon/2$, $\sigma=\pm 1$,
$\Delta\epsilon\equiv g\mu_{\rm B}H$, where $g$ is the $g$-factor, $\mu_{\rm B}$ is the Bohr magneton and $H$ is
the magnetic field. The QD Hamiltonian is thus of the following form:
\begin{equation}
\hat{H}_{\rm QD}=\sum_\sigma\epsilon_\sigma\hat{n}_\sigma+U\hat{n}_\uparrow\hat{n}_\downarrow,
\label{QD_ham}
\end{equation}
where the number operator $\hat{n}_\sigma$ is given in terms of the original Anderson fermionic operators as
\begin{equation}
\hat{n}_\sigma=d^\dagger_\sigma d_\sigma,\quad \{d_\sigma ,d^\dagger_{\sigma'}\}=\delta_{\sigma,\sigma'}\quad \{d_\sigma ,d_{\sigma'}\}=0.
\label{orig_el_op}
\end{equation}

In addition to the interacting electronic system described by the SIAM, we also consider a noninteracting
electronic system which models the contacts, that is an external system used to probe the QD. We consider two
contacts and the corresponding Hamiltonian is
\begin{equation}
\eqalign{\hat{H}_{\rm C}=\sum_{k,\sigma,x}\epsilon_{k\sigma}c^\dagger_{k\sigma x}c_{k\sigma x},\cr
\{c_{k\sigma x},c^\dagger_{k'\sigma' x'}\}=\delta_{k\sigma x,k'\sigma'x'},\quad\{c_{k\sigma x},c_{k'\sigma' x'}\}=0,}
\label{C_ham}
\end{equation}
where $k$ is the set of orbital quantum numbers in the contacts, $\sigma$ is the contact spin degree of freedom
and $x={\rm L, R}$ denotes the left and right contacts. Here we also assume that the electronic states in the
left and right contacts are characterized by the same complete set of quantum numbers, $\{k,\sigma\}$. The left
and right contacts are in equilibrium states specified by the chemical potentials $\mu_{\rm L}$ and $\mu_{\rm R}$,
respectively.

The electrons can tunnel between the QD and contacts. These tunneling events are described in terms of the
tunneling Hamiltonian,
\begin{equation}
\hat{H}_{\rm T}=\sum_{k,\sigma,x}(T_{k\sigma}c^\dagger_{k\sigma x}d_\sigma+T^{*}_{k\sigma}d^\dagger_\sigma c_{k\sigma x}),
\label{T_ham}
\end{equation}
where $T_{k\sigma}$ are the tunneling matrix elements and we assume that these matrix elements are diagonal in the
spin space and that they do not depend on $x$, that is, they are the same for the left and right contacts (QD
symmetrically coupled to the contacts).

In equilibrium the chemical potentials of the QD and contacts are equal to each other. We denote this equilibrium
value as $\mu_0$ and define the applied bias voltage $V$ as $\mu_{\rm L,R}=\mu_0\mp eV/2$.

As mentioned in the introduction, the Kondo effect, or the zero bias anomaly, develops at low temperatures out of
the Coulomb blockade regime when the QD is blockaded with an odd number of electrons. Within the SIAM this means
that the Kondo resonance takes place when the QD has one electron. This is achieved when the electron-electron
interaction $U$ exceeds the energy $\Gamma$ (see below) resulting from the QD-contacts coupling.

To describe the relevant physics and at the same time to simplify the formalism, we consider the case of of the
strongly interacting QD where $U\rightarrow\infty$. In this case the double occupancy is forbidden and the QD may
have zero or one electron. At the same time the difference $\mu_0-\epsilon_{\rm d}$ is assumed to be finite but
much larger than $\Gamma$ so that the number of electrons in the QD is close to one. Note that this asymmetric
Anderson model \cite{Hewson_1997} differs from the symmetric one where one first assumes
$\mu_0-\epsilon_{\rm d}=U/2$ and only after that takes the limit $U\rightarrow\infty$ which, as a result, leads to
$\mu_0-\epsilon_{\rm d}\rightarrow\infty$. This model admits the so-called slave-bosonic representation
\cite{Hewson_1997,Coleman_1984,Coleman_1987}. It represents a transformation where the original QD fermionic
operators $d_\sigma$, $d^\dagger_\sigma$ are mapped onto new fermionic operators $f_\sigma$, $f^\dagger_\sigma$ and bosonic
(often called slave-bosonic) operators $b$, $b^\dagger$ by means of the following relations:
\begin{equation}
\eqalign{d_\sigma=f_\sigma b^\dagger,\quad d^\dagger_\sigma=f^\dagger_\sigma b,\cr
\{f_\sigma,f^\dagger_{\sigma'}\}=\delta_{\sigma,\sigma'},\,\{f_\sigma,f_{\sigma'}\}=0,\cr
[b,b^\dagger]=1,\,[b,b]=0.}
\label{SB_transform}
\end{equation}
Additionally, the operators $f_\sigma$ and $f^\dagger_\sigma$ commute with the operators $b$, $b^\dagger$.

Physically the transformation (\ref{SB_transform}) means that instead of using the states of the electrons in the
QD one uses the states of the QD itself: the empty state ($b$, $b^\dagger$) and the state with one electron
($f_\sigma$, $f^\dagger_\sigma$).

Since the empty state and the state with one electron represent all the states of the QD, one arrives to the
constraint,
\begin{equation}
f^\dagger_\sigma f_\sigma+b^\dagger b=\hat{1},
\label{SB_constraint}
\end{equation}
which physically means that the total number of the new fermions and slave-bosons is restricted to be equal
to one.

After the slave-bosonic transformation the QD and tunneling Hamiltonians take the form \cite{Hewson_1997}:
\begin{equation}
\hat{H}_{\rm QD}=\sum_\sigma\epsilon_\sigma f^\dagger_\sigma f_\sigma,
\label{QD_ham_SB}
\end{equation}
\begin{equation}
\hat{H}_{\rm T}=\sum_{k,\sigma,x}(T_{k\sigma}c^\dagger_{k\sigma x}f_\sigma b^\dagger+T^{*}_{k\sigma}f^\dagger_\sigma b c_{k\sigma x}).
\label{T_ham_SB}
\end{equation}

The total Hamiltonian of this, in general nonequilibrium, problem is the sum of $\hat{H}_{\rm QD}$,
$\hat{H}_{\rm C}$ and $\hat{H}_{\rm T}$ given by Eqs. (\ref{QD_ham_SB}), (\ref{C_ham}) and (\ref{T_ham_SB}),
respectively. This Hamiltonian together with the constraint in Eq. (\ref{SB_constraint}) represent the theoretical
model able to describe the essential behavior of the zero bias anomaly arising due to the Kondo effect in the
presence of an external magnetic field.
\section{Keldysh field integral solution and its asymptotics}\label{KFIS}
As it is well known, the Kondo effect is an essentially nonperturbative phenomenon \cite{Hewson_1997}. Therefore,
to cover all the regimes of the Kondo state one must resort to nonperturbative methods able to deal with both
electron-electron interactions and nonequilibrium.

In particular, the real-time formalism developed by Keldysh \cite{Keldysh_1965} represents a general and powerful
method to treat interacting many-particle systems both in equilibrium and nonequilibrium. Originally this method
was developed in the diagrammatic framework. Its field integral form has been developed in Ref. \cite{Kamenev_1999}
to study the interplay between disorder and electron-electron interactions in metals. The advantage of the field
integral form is that it reduces the problem to an analysis of a functional, the so-called Keldysh effective action.
This analysis has a more systematic nonperturbative character than the diagrammatic selection on the basis of the
topological structure reflecting the physical content of a given diagram.

In addition to being systematic, this nonperturbative approach turns out to be very general and applicable also to
mesoscopic and QD systems. It has been applied, {\it e.g.}, to describe the Coulomb blockade in QDs
\cite{Altland_2009,Altland_2010} as well as the Kondo effect in the strong coupling limit \cite{Ratiani_2009}, weak
coupling limit \cite{Smirnov_2011,Smirnov_2011a} and in the whole energy range \cite{Smirnov_2013}.

In particular, the slave-bosonic formulation presented in the previous section has been used in the Keldysh effective
action framework in Refs. \cite{Smirnov_2011,Smirnov_2011a,Smirnov_2013}. The advantage of these slave-bosonic
theories over slave-boson mean-field theories \cite{Hewson_1997,Lopez_2003} and $1/N$ expansions \cite{Ratiani_2009}
is that they take into account the constraint in Eq. (\ref{SB_constraint}) exactly while slave-boson mean-field
theories and $1/N$ expansions account for this constraint only approximately. Because of this approximation one can
only access the low energy (smaller than $kT_{\rm K}$) physics of the Kondo state, that is the Fermi-liquid regime,
or strong coupling limit, characterized by the quadratic dependence of the QD differential conductance. The main
drawback of this approximation is that it does not lead to logarithmic terms \cite{Hewson_1997} in the differential
conductance and thus one cannot access intermediate (of order of $kT_{\rm K}$) and high energy (larger than
$kT_{\rm K}$) physics, that is, the crossover region and weak coupling limit of the Kondo state, respectively. In
contrast, the slave-bosonic Keldysh effective action theories in Refs.
\cite{Smirnov_2011,Smirnov_2011a,Smirnov_2013} account for the constraint in Eq. (\ref{SB_constraint}) exactly
and thus contain logarithmic terms.
\begin{figure}
\centering
\includegraphics[width=8.0 cm]{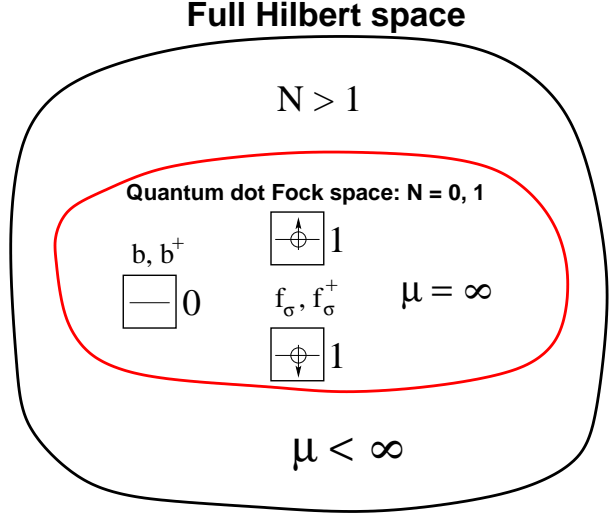}
\caption{\label{figure_1} To formulate the field integral in Eq. (\ref{KFI_observ}) the constraint in
  Eq. (\ref{SB_constraint}) is temporarily removed by introducing a positive real parameter $\mu$. When $\mu$ is
  finite, $0<\mu<\infty$, the QD may have any number of the new fermions and slave-bosons. One projects onto the
  physical subspace, the QD Fock space, by taking the limit $\mu\rightarrow\infty$ after the field integration has
  been performed.}
\end{figure}

As shown below, the formalism of Ref. \cite{Smirnov_2013} is straightforwardly generalized to the case of a QD
in an external magnetic field.

The theory of Ref. \cite{Smirnov_2013} introduces a complex function $E_\alpha$ which is found from the condition
that the QD TDOS,
\begin{equation}
\nu_\sigma(\epsilon)\equiv-\frac{1}{\pi\hbar}{\rm Im}[G^+_{\sigma\sigma}(\epsilon)],
\label{TDOS_def}
\end{equation}
where $G^+_{\sigma\sigma'}(\epsilon)$ is the QD retarded Green's function, in equilibrium and at zero temperature has a peak
at $\epsilon=\mu_0$ and the amplitude of this peak is equal to the unitary limit.

In the case of a QD in an external magnetic field, one obtains the Keldysh effective action $S_{\rm eff}$ using a
derivation similar to the one in Ref. \cite{Smirnov_2013} and expresses any physical observable
$\hat{O}=\mathcal{F}(d_\sigma^\dagger$, $d_\sigma)$ as
\begin{equation}
\eqalign{\langle\hat{O}\rangle(t)=\frac{1}{\mathcal{N}_0}\underset{\mu\rightarrow\infty}{\lim}e^{\beta\mu}
\!\!\int\!\!\mathcal{D}[\bar{\chi},\chi]e^{\frac{i}{\hbar}S_{\rm eff}[\bar{\chi}^{\rm cl,q}(\tilde{t});\chi^{\rm cl,q}(\tilde{t})]}
\mathcal{F}[\bar{\chi}^{\rm cl,q}(t);\chi^{\rm cl,q}(t)],\cr
S_{\rm eff}[\bar{\chi}^{\rm cl,q}(t);\chi^{\rm cl,q}(t)]=S_0[\bar{\chi}^{\rm cl,q}(t);\chi^{\rm cl,q}(t)]+
S_{\rm T}[\bar{\chi}^{\rm cl,q}(t);\chi^{\rm cl,q}(t)],}
\label{KFI_observ}
\end{equation}
where $\chi^{\rm cl,q}$ are the classical and quantum \cite{Altland_2010} eigenstates of the bosonic annihilation
operator $b$, $\beta\equiv 1/kT$ is the inverse temperature and $\mathcal{N}_0$ is a normalization constant
\cite{Smirnov_2011}. The limit $\mu\rightarrow\infty$ in Eq. (\ref{KFI_observ}) takes into account the constraint in
Eq. (\ref{SB_constraint}). The point is that the field integral in Eq. (\ref{KFI_observ}) is most simply formulated for
a system with arbitrary numbers of the new fermions and slave-bosons, that is temporarily removing the constraint in Eq.
(\ref{SB_constraint}). One achieves such a formulation introducing a real positive parameter $\mu$. As shown in Fig.
\ref{figure_1}, for $0<\mu<\infty$ the QD is allowed to have arbitrary numbers of the new fermions and slave-bosons,
that is one deals with the full Hilbert space. One projects onto the physical QD Fock space by taking the limit
$\mu\rightarrow\infty$.

In Eq. (\ref{KFI_observ}) $S_0[\bar{\chi}^{\rm cl,q}(t);\chi^{\rm cl,q}(t)]$ is the standard free bosonic action
\cite{Altland_2010} on the Keldysh contour and
$S_{\rm T}[\bar{\chi}^{\rm cl,q}(t);\chi^{\rm cl,q}(t)]=-i\hbar\,{\rm tr}\ln\bigl[-iG^{(0)-1}-i\mathcal{T}\bigl]$ is the
tunneling action of the problem. Here the matrix $\mathcal{T}$ is off-diagonal in the QD-contacts space,
\begin{equation}
\mathcal{T}=
\left(
\matrix{
0 & M_{\rm T}^\dagger(\sigma t|k'\sigma't') \cr
M_{\rm T}(k\sigma t|\sigma't') & 0
}
\right),
\label{T_matr}
\end{equation}
\begin{equation}
M_{\rm T}(k\sigma t|\sigma' t')=\frac{\delta(t-t')\delta_{\sigma\sigma'}T_{k\sigma}}{\sqrt{2}\hbar}
\left(\matrix{
\bar{\chi}^{\rm cl}(t)-\gamma_\sigma\sqrt{2} & \bar{\chi}^{\rm q}(t) \cr
\bar{\chi}^{\rm q}(t) & \bar{\chi}^{\rm cl}(t)-\gamma_\sigma\sqrt{2}}\right).
\label{tun_matr}
\end{equation}
The Green's function matrix $G^{(0)}$ is block-diagonal in the QD-contacts space. Its QD block $G^{(0)}_{\rm d}$ has the
standard $2\times 2$ fermionic Keldysh structure:
\begin{equation}
G^{(0)}_{\rm d}(\sigma t|\sigma' t')=\delta_{\sigma\sigma'}
\left(\matrix{
G^+_{\sigma}(t-t') & G^{\rm K}_{\sigma}(t-t') \cr
0 & G^-_{\sigma}(t-t')}\right).
\label{G_0_matr}
\end{equation}
In the frequency domain the components of the above matrix are
\begin{equation}
\eqalign{G^+_\sigma(\omega)=\frac{\hbar}{\hbar\omega-(\epsilon_\sigma+\mu)+iE_{\alpha,\sigma}},\quad
G^-_\sigma(\omega)=[G^+_\sigma(\omega)]^*,\cr
G^{\rm K}_\sigma(\omega)=\frac{1}{2}[G^+_\sigma(\omega)-G^-_\sigma(\omega)]\sum_x\tanh\biggl[\frac{\hbar\omega-\mu_x}{2kT}\biggl].}
\label{G_0_matr_como_fr}
\end{equation}
Here $E_{\alpha,\sigma}\equiv\alpha_\sigma\Gamma/2$, $\alpha_\sigma\equiv\gamma_\sigma\delta_\sigma$,
$\Gamma\equiv 2\pi\nu_{\rm C}|\tau|^2$, where $\nu_{\rm C}$ characterizes the contacts Lorentzian density of sates,
$\sum_k\delta(\epsilon-\epsilon_{k\sigma})=(\nu_{\rm C}W^2)/2(\epsilon^2+W^2)$, of width $2W$ and $\tau$ is the value of the
tunneling matrix element $T_{k\sigma}$ assumed to be independent of $k$, $\sigma$.

In the case of the QD TDOS for spin $\sigma$ the expression for the integrand in Eq. (\ref{KFI_observ}) is
\begin{equation}
\eqalign{\mathcal{F}[\bar{\chi}^{\rm cl,q}(t);\chi^{\rm cl,q}(t)]=[\bar{\chi}_-(t)\chi_+(0)-\bar{\chi}_+(t)\chi_-(0)]\times\cr
\times[G^{(0)-1}+\mathcal{T}]^{-1}(\sigma t|\sigma 0),}
\label{F_TDOS}
\end{equation}
where $\bar{\chi}_{\pm}$, $\chi_{\pm}$ are the slave-bosonic fields on the forward and backward branches
\cite{Altland_2010} of the Keldysh contour.

Anticipating that the Kondo physics arises from the spin-flip processes, to obtain the QD TDOS for fixed spin $\sigma$
within the lowest order expansion (see below), we first choose in the $\sigma$- and $-\sigma$-blocks of the matrix
$G^{(0)}_{\rm d}$ the imaginary parts of $E_{\alpha,\sigma}$ and $E_{\alpha,-\sigma}$ as
$E_{\alpha,\sigma}^{\rm I}=E_\alpha^{\rm I}-\sigma\Delta\epsilon$ and $E_{\alpha,-\sigma}^{\rm I}=E_\alpha^{\rm I}$, while the real
parts are $E_{\alpha,\sigma}^{\rm R}=E_{\alpha,-\sigma}^{\rm R}=E_\alpha^{\rm R}$. Here $E_\alpha$ is found, as in
Ref. \cite{Smirnov_2013}, at zero magnetic field, yielding $E^{\rm R}_\alpha/kT_{\rm K}\approx 1.421416$,
$E^{\rm I}_\alpha/kT_{\rm K}\approx 1.051215$ with $kT_{\rm K}=2W\exp[-\pi(\mu_0-\epsilon_{\rm d})/\Gamma]$. After this choice the
$\sigma$- and $-\sigma$-blocks of $G^{(0)}_{\rm d}$ are both expressed through the spin $-\sigma$. Finally, as in Ref.
\cite{Smirnov_2013}, we expand $S_{\rm eff}$ and $\mathcal{F}$ up to the second order in the slave-bosonic fields in
$\chi^{\rm cl}(t)-\delta_\sigma\sqrt{2}$, $\bar{\chi}^{\rm cl}(t)-\gamma_\sigma\sqrt{2}$ and $\chi^{\rm q}(t)$, $\bar{\chi}^{\rm q}(t)$.

The resulting QD TDOS obtained after performing the Gaussian field integral in Eq. (\ref{KFI_observ}) is then given by
an expression similar to the one in Ref. \cite{Smirnov_2013} except for the spin dependence arising due to the
Zeeman splitting,
\begin{equation}
\nu_\sigma(\epsilon)=
\frac{1}{2\pi}\frac{\Gamma}{[\epsilon_{\rm d}+\sigma\Delta\epsilon/2-\epsilon+\Gamma\Sigma_{\sigma{\rm R}}^+(\epsilon)]^2+[\Gamma\Sigma_{\sigma{\rm I}}^+(\epsilon)]^2},
\label{TDOS_result}
\end{equation}
\begin{equation}
\eqalign{\Sigma^+_\sigma(\epsilon)=
\sum_x\biggl\{\frac{1}{4\pi}\psi\biggl[\frac{1}{2}-\frac{W}{2\pi kT}\biggl]+\frac{1}{4\pi}\psi\biggl[\frac{1}{2}+\frac{W}{2\pi kT}\biggl]-\cr
-\frac{1}{2\pi}\psi\biggl[\frac{1}{2}+\frac{E_\alpha}{2\pi kT}-\frac{i\mu_x}{2\pi kT}+\frac{i(\epsilon-\sigma\Delta\epsilon)}{2\pi kT}\biggl]+
\frac{i}{2}\frac{1}{\exp\bigl(\frac{-\mu_x+iW}{kT}\bigl)+1}\biggl\},}
\label{Rtrd_se}
\end{equation}
where $\psi$ is the digamma function.

Here we would like to mention the main drawback of the approximations made in deriving Eqs. (\ref{TDOS_result}) and
(\ref{Rtrd_se}). Due to the second order expansion of Eq. (\ref{F_TDOS}), the theory does not take into account all
inelastic cotunneling processes which are of the fourth order in the tunneling matrix elements. These processes are
included only partly, through the effective action. Therefore, the present theory underestimates
$\nu_\uparrow(\epsilon)$ at energies $\epsilon-\mu_0>g\mu_{\rm B}H$ ($H>0$) and $\nu_\downarrow(\epsilon)$ at energies
$\epsilon-\mu_0<-g\mu_{\rm B}H$. Furthermore, it is known \cite{Moore_2000,Rosch_2003} that for the $s-d$ model at very
small magnetic fields ($g\mu_{\rm B}|H|\ll kT_{\rm K}$) the QD TDOS peak is located at $\epsilon=(2/3)g\mu_{\rm B}H$ and
at $\epsilon=g\mu_{\rm B}H$ at larger fields. Our theory for the highly asymmetric SIAM predicts, as one can see from
Eqs. (\ref{TDOS_result}) and (\ref{Rtrd_se}), that the peak is always located close to $\epsilon=g\mu_{\rm B}H$. This
perhaps can be attributed partly to the quality of our approximation and partly to the physical difference between the
$s-d$ model and the highly asymmetric SIAM. It is technically complicated to include higher order terms in Eq.
(\ref{F_TDOS}), and the corresponding theory will be the focus of our future research. However, below we demonstrate
that already this simple theory has a number of advantages and is in a good agreement with some other theoretical
predictions as well as with experiments.

With the QD TDOS (\ref{TDOS_result}) we can calculate the differential conductance using the expression
\cite{Meir_1993,Wingreen_1994} for the current through the QD (see, for example, Eqs. (3) and (29) in Ref.
\cite{Wingreen_1994}),
\begin{equation}
\eqalign{I=\frac{e}{\hbar}\sum_\sigma\int_{-\infty}^\infty \!\!\!\!d\epsilon[n_{{\rm R}}(\epsilon)-n_{{\rm L}}(\epsilon)]\frac{\Gamma}{4}\frac{W^2}{\epsilon^2+W^2}\nu_\sigma(\epsilon),\cr
n_{{\rm L,R}}(\epsilon)=\frac{1}{\exp[\beta(\epsilon-\mu_0\pm eV/2)]+1}.}
\label{MW_expr}
\end{equation}

The differential conductance $\sigma_d$ is obtained from Eq. (\ref{MW_expr}) via the derivative of the current through
the QD with respect to the bias voltage, $\sigma_d=\partial I/\partial V$. The differential conductance is in general
a function of the temperature, voltage and magnetic field, $\sigma_d=\sigma_d(T,V,H)$. It can be obtained substituting
the QD TDOS, Eq. (\ref{TDOS_result}), into the expression for the current through the QD, Eq. (\ref{MW_expr}). This is
a difficult task because the QD TDOS (\ref{TDOS_result}) is already a complicated function which, in addition, must be
integrated. Therefore, in general one has to perform a numerical integration to obtain the differential conductance.
\begin{figure}
\centering
\includegraphics[width=9.0 cm]{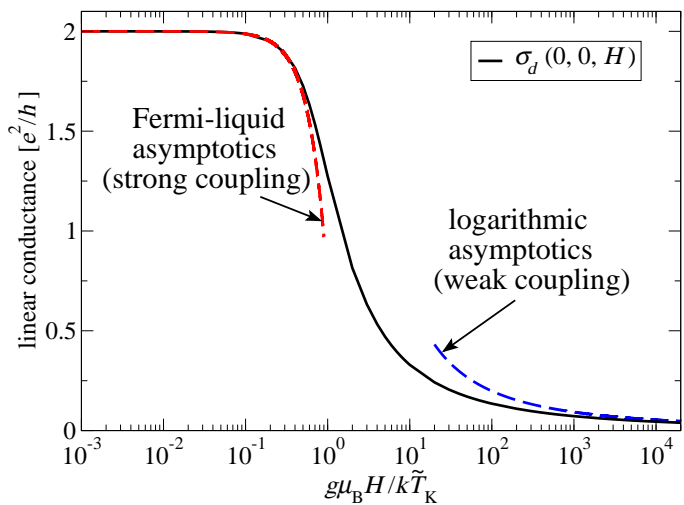}
\caption{\label{figure_2} The QD universal differential conductance as a function of the magnetic field
in the Kondo regime at $T=0$ and $V=0$. We use $\mu_0-\epsilon_{\rm d}=7\,\Gamma$, $W=100\,\Gamma$ which gives
$kT_{\rm K}/\Gamma\approx 5.63\cdot 10^{-8}$. The Kondo temperature $\widetilde{T}_{\rm K}$ is defined in the text.
The black solid curve shows the differential conductance obtained from Eqs. (\ref{TDOS_result}), (\ref{Rtrd_se}) and
(\ref{MW_expr}). The dashed lines show the Fermi-liquid (red/upper) and logarithmic (blue/lower) asymptotics given
by Eqs. (\ref{DC_le}) and (\ref{LC_hmf}), respectively.}
\end{figure}

However, at low energies it turns out to be possible to obtain the differential conductance analytically. From Eqs.
(\ref{TDOS_result}), (\ref{Rtrd_se}) and (\ref{MW_expr}) it follows that at $T<T_{\rm K}$, $e|V|<kT_{\rm K}$ and
$g\mu_{\rm B}|H|<kT_{\rm K}$ the differential conductance takes the form:
\begin{equation}
\sigma_d(T,V,H)=\frac{2e^2}{h}\biggl[1-c_T\biggl(\frac{T}{T_{\rm K}}\biggl)^2-c_V\biggl(\frac{eV}{kT_{\rm K}}\biggl)^2-
c_H\biggl(\frac{g\mu_{\rm B}H}{kT_{\rm K}}\biggl)^2\biggl],
\label{DC_le}
\end{equation}
which shows that at low energies the QD TDOS (\ref{TDOS_result}) leads to two important consequences. First, the
differential conductance is a universal function of the temperature, bias voltage and magnetic field with the correct
scaling given by the Kondo temperature $T_{\rm K}$ of the corresponding problem without magnetic field. Second, the
differential conductance does not contain terms linear in $T$, $V$ and $H$, which means that the QD TDOS
(\ref{TDOS_result}) correctly reflects the low energy physics leading to the Fermi-liquid behavior. Here it is
important to note that linear in $V$ terms are absent due to the symmetric coupling of the QD to the contacts. Had one
assumed that the couplings of the QD to the right and left contacts are different, linear in $V$ terms would have had
to appear \cite{Aligia_2012,Munoz_2013}.

The Fermi-liquid coefficients are obtained from a Taylor expansion of $\sigma_d(T,V,H)$ similar to the one used in Ref.
\cite{Smirnov_2013},
\begin{equation}
\eqalign{c_T=\frac{4}{3}\frac{2\ln(2|\mathcal{E}_\alpha|)+1}{|\mathcal{E}_\alpha|^2},\quad c_V=\frac{1}{\pi^2}\frac{4\ln(2|\mathcal{E}_\alpha|)+1}{|\mathcal{E}_\alpha|^2},\cr
c_H=\frac{4}{\pi^2}\frac{\ln(2|\mathcal{E}_\alpha|)+1}{|\mathcal{E}_\alpha|^2},}
\label{FL_coeff}
\end{equation}
with the following universal ratios
\begin{equation}
\eqalign{\frac{c_V}{c_T}=\frac{3}{2\pi^2}\frac{4\ln(2|\mathcal{E}_\alpha|)+1}{4\ln(2|\mathcal{E}_\alpha|)+2}\approx 0.13043,\cr
\frac{c_H}{c_T}=\frac{3}{2\pi^2}\frac{2\ln(2|\mathcal{E}_\alpha|)+2}{2\ln(2|\mathcal{E}_\alpha|)+1}\approx 0.19509,}
\label{FL_coeff_ratio}
\end{equation}
where the complex universal ratio $E_\alpha/kT_{\rm K}$ is denoted through $\mathcal{E}_\alpha$,
$\mathcal{E}_\alpha\equiv E_\alpha/kT_{\rm K}$, with the value specified above (see the text before Eq. (\ref{TDOS_result})).
For comparison we also give the universal ratios \cite{Oguri_2001,Oguri_2005,Merker_2013} for the symmetric Anderson
model, where one first puts $\mu_0-\epsilon_d=U/2$ and only after that takes the limit
$U\rightarrow\infty$: $c_V/c_T=3/(2\pi^2)\approx 0.15198$, $c_H/c_T=1/\pi^2\approx 0.10132$.

In equilibrium and at zero temperature one obtains from Eqs. (\ref{TDOS_result}), (\ref{Rtrd_se}) and (\ref{MW_expr})
the differential conductance at $g\mu_{\rm B}|H|\gg kT_{\rm K}$:
\begin{equation}
\sigma_d(0,0,H)=\frac{2e^2}{h}\frac{\pi^2}{4}\frac{1}{\ln^2\bigl(\frac{g\mu_{\rm B}|H|}{kT_{\rm K}}\bigl)}.
\label{LC_hmf}
\end{equation}
As one can see, at large magnetic fields the differential conductance at $T=0$ and $V=0$ is also a universal function
of the magnetic field with the correct scaling. Moreover, it correctly reflects the high energy physics leading to the
logarithmic behavior \cite{Hewson_1997}.
\begin{figure}
\centering
\includegraphics[width=9.0 cm]{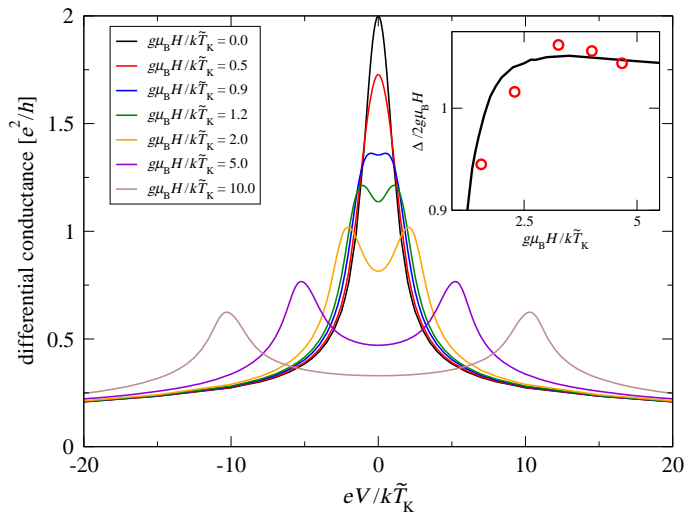}
\caption{\label{figure_3} The universal differential conductance as a function of the bias voltage at
zero temperature and different magnetic fields or Zeeman splittings $g\mu_{\rm B}H$. The parameters are the same as for
Fig. \ref{figure_2}. When the magnetic field increases the zero bias anomaly first decreases and then starts to split
into two peaks. This happens at the critical magnetic field $g\mu_{\rm B}H_c\approx 0.8\,k\widetilde{T}_{\rm K}$. The
inset shows the distance $\Delta$ between these two peaks as a function of the magnetic field. The circles in the
inset show the experimental data \cite{Quay_2007}.}
\end{figure}

Already these low- and high-energy asymptotics demonstrate advantages of the present theory over other slave-bosonic
theories like mean-field theories or NCA which do not have the logarithmic or Fermi-liquid behavior, respectively
\cite{Hewson_1997}.

In the next section we obtain the differential conductance from Eqs. (\ref{TDOS_result}), (\ref{Rtrd_se}) and
(\ref{MW_expr}) in the whole energy range both in equilibrium and nonequilibrium. However, to obtain it in this case
one has, in general, to perform the integration in Eq. (\ref{MW_expr}) numerically.
\section{Universal results in the whole energy range}\label{Results}
It has been demonstrated in Ref. \cite{Smirnov_2013} that the differential conductance resulting from the
Keldysh effective action theory is a universal function of the temperature and bias voltage in the whole energy range.
It has also been shown that it has the correct scaling given by the Kondo temperature $T_{\rm K}$. Here we provide
the differential conductance resulting from the Keldysh effective action theory taking into account an external
magnetic field, Eqs. (\ref{TDOS_result}) and (\ref{Rtrd_se}). Using Eq. (\ref{MW_expr}), we obtain the linear
conductance as a function of the magnetic field at $T=0$, $\sigma_d(0,0,H)$. It is shown in Fig. \ref{figure_2}. As one
expects, in the Kondo regime $\sigma_d(0,0,H)$ turns out to be a universal function of $g\mu_{\rm B}H/kT_{\rm K}$ (or
$g\mu_{\rm B}H/k\widetilde{T}_{\rm K}$, where $\widetilde{T}_{\rm K}$ is defined as
$\sigma_d(\widetilde{T}_{\rm K},0,0)=\sigma_d(0,0,0)/2$, $\widetilde{T}_{\rm K}\approx 1.47\,T_{\rm K}$).

Let us now address the nonequilibrium Kondo physics in the presence of an external magnetic field. In Fig.
\ref{figure_3} we show the zero temperature differential conductance as a function of the bias voltage at different
magnetic fields. Again all the curves are universal with the scaling given by the Kondo temperature, which means that
Eqs. (\ref{TDOS_result}) and (\ref{Rtrd_se}) provide the differential conductance with the correct scaling also in
nonequilibrium and finite magnetic fields. From Fig. \ref{figure_3} one observes that when the magnetic field is
increased, the zero bias Kondo peak first decreases and then splits into two peaks. Our theory predicts that for the
asymmetric SIAM, considered in the present work, the zero bias peak starts to split at the critical magnetic field
$g\mu_{\rm B}H_c\approx 0.8\,k\widetilde{T}_{\rm K}$ which is close to other theoretical results
\cite{Hewson_2005,Aligia_2006}, $g\mu_{\rm B}H_c\approx 0.6\,k\widetilde{T}_{\rm K}$,
\begin{figure}
\centering
\includegraphics[width=9.0 cm]{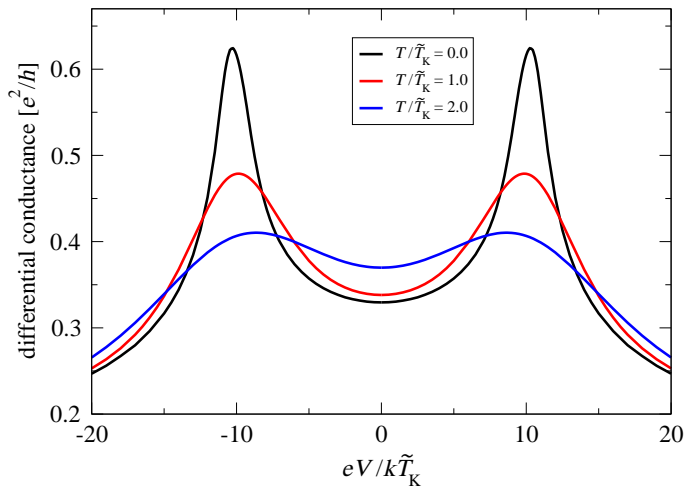}
\caption{\label{figure_4} The universal differential conductance as a function of the bias voltage at
  finite temperatures and magnetic field $g\mu_{\rm B}H=10\,kT_{\rm K}$. The parameters are the same as for Fig.
  \ref{figure_2}. The increase in temperature leads to two effects: 1) the height of the peaks reduces; 2) the width of
  the peaks increases (broadening).}
\end{figure}
$g\mu_{\rm B}H_c\approx 0.7\,k\widetilde{T}_{\rm K}$ obtained, however, for the symmetric SIAM. It is interesting to note
that an equilibrium estimate \cite{Costi_2000} predicts $g\mu_{\rm B}H_c\approx 1.0\,k\widetilde{T}_{\rm K}$ for the case
of the symmetric SIAM. This shows that the aspect of the critical magnetic field requires a proper nonequilibrium
treatment. For example, a qualitative treatment of this aspect with the help of the equation of motion method
\cite{Dong_2001} significantly overestimates the critical magnetic field and predicts
$g\mu_{\rm B}H_c\approx 2.67\,k\widetilde{T}_{\rm K}$. The inset of Fig. \ref{figure_3} shows the distance $\Delta$
between the two peaks as a function of the magnetic field. At low magnetic fields $\Delta$ has a rapid increase and
then saturates at $\Delta/2g\mu_{\rm B}H\approx 1.017$. This high field limit of our theory is in a good agreement
with equilibrium \cite{Moore_2000,Zitko_2009} and nonequilibrium \cite{Schmitt_2011} theoretical results which predict
that in the Kondo regime at high magnetic fields $\Delta$ saturates at a value close to two Zeeman splittings,
$2g\mu_{\rm B}H$. Moreover, in the Kondo regime (that is at not too large magnetic fields so that the charge excitations
are not involved in the transport) the differential conductance obtained in Ref. \cite{Schmitt_2011}, using a
nonequilibrium extension of NRG, is in a reasonable agreement with our results (the red curve in Fig. 2b of Ref.
\cite{Schmitt_2011} and the curve for $g\mu_{\rm B}H=5.0\,k\widetilde{T}_{\rm K}$ in our Fig. \ref{figure_3}). As one can
see from the inset of Fig. \ref{figure_3}, our nonequilibrium theory is also in a good agreement with experiments in
Ref. \cite{Quay_2007}. In contrast, the equilibrium theories in Refs. \cite{Costi_2000,Moore_2000} both predict a
saturation at much larger fields than observed in Ref. \cite{Quay_2007}. At the same time, similar to our
nonequilibrium theory, a nonequilibrium extension of NRG \cite{Schmitt_2011} predicts for the symmetric SIAM with
finite $U$ a faster increase of $\Delta$ as a function of the magnetic field (see Figs. 3b and 3c in Ref.
\cite{Schmitt_2011}).

Finally, in Fig. \ref{figure_4} we show the differential conductance as a function of the bias voltage at finite
temperatures and at the magnetic field value $g\mu_{\rm B}H=10\,k\widetilde{T}_{\rm K}$. Also in this case the
differential conductance is a universal function with correct scaling given by the Kondo temperature. As one
expects, when the temperature increases the two peaks in the differential conductance are lowered and broadened.
\section{Conclusion}\label{Conclusion}
In conclusion, we have applied the Keldysh effective action theory to study the Kondo effect in QDs in an external
magnetic field both in equilibrium and nonequilibrium. To this end we have generalized the Keldysh effective action
theory from Ref. \cite{Smirnov_2013} to take into account the effect of the magnetic field through the Zeeman
splitting. An approximate analytical expression for the QD TDOS has been obtained. We have used this QD TDOS to
calculate the differential conductance in the whole energy range and demonstrated that the theory has a number of
advantages: 1) Fermi-liquid behavior when the Zeeman energy is much less than $kT_{\rm K}$ (advantage over NCA
\cite{Hewson_1997}); 2) logarithmic behavior when the Zeeman energy is much large than $kT_{\rm K}$ (advantage over
mean-field theories \cite{Hewson_1997}); 3) critical magnetic field close to results of previous theories
\cite{Hewson_2005,Aligia_2006} and experiments \cite{Quay_2007}; 4) distance between the maxima of the differential
conductance as a function of the magnetic field in a good agreement with experiments \cite{Quay_2007} and absence of
a spurious zero bias peak at a finite magnetic field (advantage over NCA \cite{Meir_1993}); 5) high-field limit of
this distance in a good agreement with experiments \cite{Quay_2007}; 6) universality with the correct scaling given
by $T_{\rm K}$.

At the same time, due to the second order expansion of Eq. (\ref{F_TDOS}), the theory does not take into account
all inelastic cotunneling processes which are of the fourth order in the tunneling matrix elements. Therefore, the
present theory underestimates the differential conductance at voltages such that $e|V|>g\mu_{\rm B}|H|$ (see Fig.
\ref{figure_3}). An additional consequence of this approximation is that at high magnetic fields the differential
conductance as a function of the bias voltage is oversensitive with respect to the temperature (see Fig.
\ref{figure_4}). These drawbacks can be eliminated by taking into account higher order terms in Eq. (\ref{F_TDOS}).
This is technically more complicated and will be addressed in our future research.
\section*{Acknowledgments}
The authors thank Theodoulos Costi for a valuable discussion. Support from the DFG under the program SFB 689 is
acknowledged.
\section*{References}
\bibliographystyle{iopart-num}
\providecommand{\newblock}{}

\end{document}